\documentclass[aps,prd,twocolumn,floats,floatfix,showpacs,superscriptaddress,nofootinbib]{revtex4-2}

\usepackage{graphicx,amssymb,amsmath,amsthm,amsfonts,epsfig,epsf,fixmath}
\usepackage[usenames]{color}
\usepackage{epstopdf}
\usepackage{booktabs}
\usepackage{aas_macros}
\usepackage{bm}
\usepackage{dcolumn}
\usepackage[dvipsnames]{xcolor}
\usepackage[colorlinks]{hyperref}
\usepackage{cleveref}
\usepackage[utf8]{inputenc}
\usepackage{resizegather}

\begin{document}

\title{Quasi-normal modes of rotating black holes in Einstein-dilaton Gauss-Bonnet gravity: the first order in rotation}

\author{Lorenzo Pierini}
	\email{lorenzo.pierini@uniroma1.it}
	\affiliation{Dipartimento di Fisica, ``Sapienza" Università di Roma \& Sezione INFN Roma1, Piazzale Aldo Moro 
5, 00185, Roma, Italy}
\author{Leonardo Gualtieri}
	\email{leonardo.gualtieri@roma1.infn.it}
	\affiliation{Dipartimento di Fisica, ``Sapienza" Università di Roma \& Sezione INFN Roma1, Piazzale Aldo Moro 
5, 00185, Roma, Italy}

\begin{abstract}
Gravitational spectroscopy - the measurement of the quasi-normal modes of a black hole from the ringdown signal of a
binary black hole coalescence - is one of the most promising tools to test gravity in the strong-field, large-curvature
regime, but without the knowledge of the black hole quasi-normal modes in specific cases of modified gravity theories,
only null tests of general relativity are possible.  More specifically, we need to know the modes of rotating black
holes, because typical compact binary mergers lead to black holes with large spins. In this article we compute, for the
first time, the gravitational quasi-normal modes of rotating black holes in a modified gravity theory, up to first order
in the spin. We consider Einstein-dilaton Gauss-Bonnet gravity, one of the simplest modifications of general relativity
in the large-curvature regime. We find that the shifts in the mode frequencies and damping times due to general
relativity modifications are significantly magnified by rotation.
\end{abstract}

\maketitle

\section{Introduction} \label{sec:intro}
The detection of gravitational waves (GWs) from a binary black hole (BH) coalescence~\cite{Abbott:2016blz} opened the
possibility of testing general relativity (GR) in the strong-field regime of
gravity~\cite{TheLIGOScientific:2016src}. While the inspiral signal is very sensitive to dipolar emission due to
additional polarizations (see e.g.~\cite{Barausse:2016eii,Yunes:2016jcc}), the ringdown signal, i.e. the emission of GWs
from the oscillating BH born in the merger, carries the imprint of possible GR modifications in the large-curvature
regime.  The merger signal may also be a very sensitive probe of strong-field GR modifications, but the necessary
theoretical modelling of this stage is extremely challenging; indeed, in order to perform numerical relativity
simulations of binary BH coalescence in a modified gravity theory, we need to develop a well-posed and well-behaved
formulation of the time-evolution problem (see
e.g.~\cite{Delsate:2014hba,Papallo:2017qvl,Ripley:2019irj,Kovacs:2020ywu,Witek:2020uzz,Julie:2020vov,East:2020hgw}).

The study of the ringwdown is also technically challenging, but it does not present fundamental issues. After the
merger, the waveform is accurately described by a superposition of damped oscillations, the quasi-normal modes (QNMs) of
the BH~\cite{Kokkotas:1999bd,Ferrari:2007dd,Berti:2009kk}. The frequencies and damping times of the modes in the
observed signal can be compared with the predictions of GR, leading to null tests of GR. If a deviation is observed, a
comparison with the predictions of specific modifed gravity theories would give information on the nature of the
modification. This idea has been called ``gravitational spectroscopy''~\cite{Dreyer:2003bv,Berti:2005ys,Berti:2007zu}
(see also \cite{Berti:2018vdi} and references therein).

Gravitational spectroscopy requires an {\it a priori} theoretical knowledge of the QNMs in modified gravity
theories. Although the computation of the QNMs is conceptually simple - it requires the numerical integration of the
field equations linearized around the stationary BH background, imposing appropriate boundary conditions at the horizon
and at infinity - few results are currently available in the literature.  The QNMs of static, non-rotating BHs have been
computed for a certain number of modified gravity
theories~\cite{Cardoso:2009pk,Molina:2010fb,Kobayashi:2012kh,Kobayashi:2014wsa,Salcedo2016,Tattersall:2019nmh,Blazquez-Salcedo:2020caw,Blazquez-Salcedo:2017txk}
(other computations refer to oscillations of a test scalar field, and thus are not relevant for gravitational
spectroscopy).  A different possible approach to study BH QNMs in modified gravity is that of modifying the radial
potential in the perturbation equations, and computing how these deformations affect the QNM frequencies and damping
times~\cite{Cardoso:2019mqo,McManus:2019ulj}.

The main limitation of these studies is that they only refer to {\it static, non-rotating} BHs.  Conversely,
astrophysical BHs {\it do} rotate; due to angular momentum conservation, even when the two BHs in the binary have
negligible spin, the final BH has --~for a comparable-mass binary- $\bar a=J/M^2\sim0.7$ (where $M,J$ are the mass and
angular momentum of the BH, respectively). Gravitational spectroscopy, then, requires the previous knowledge of QNMs of
{\it stationary, rotating} BHs in modified gravity theories. This article is, to our knowledge, the first computation of
rotational corrections of BH QNMs in a modifed theory of gravity.

Several modifiations of GR have been proposed, but most of them can be reformulated in terms of the addition of extra
polarizations - such as a scalar field - to the gravitational sector of the theory~\cite{TopicalReview}. Some of these
{\it scalar-tensor theories} modify the weak-field regime of gravity, and thus are strongly constrained by binary pulsar
observations~\cite{Antoniadis:2013pzd,TopicalReview}. This is not the case for Einstein-dilaton Gauss-Bonnet (EdGB)
gravity, one of the simplest scalar-tensor theories which modifies the strong-field, large-curvature regime of
gravity. EdGB gravity is described by the action~\cite{Mignemi:1992nt,Kanti1995}:
\begin{equation}
  S = \int d^4 x \frac{\sqrt{-g}}{16 \pi} \left(R-\frac{1}{2} \partial_\mu \phi \partial^\mu \phi
  + \frac{\alpha_{\rm GB}}{4} e^\phi \mathcal{R}^2_{\rm GB} \right)+S_{\rm m}
\label{eq:S_edgb}
\end{equation}
where
\begin{equation}
\mathcal{R}^2_{\rm GB}=R_{\mu\nu\rho\sigma} R^{\mu\nu\rho\sigma} -4 R_{\mu\nu} R^{\mu\nu} +R^2
\label{eq:gauss-bonnet-term}
\end{equation}
is the Gauss-Bonnet term, and $S_{\rm m}$ is the matter Lagrangian. This theory - which naturally arises in some
low-energy truncations of string theories - belongs to the class of Horndeski
gravity~\cite{Horndeski:1974wa,Kobayashi:2019hrl}, the scalar-tensor theories with second-order-in-time field equations
(which are thus free from the Ostrogradsky instability). Remarkably, while in several scalar-tensor theories the no-hair
theorem of GR is satisfied, and thus stationary BHs are described by the Kerr solution, BHs in EdGB gravity always have
a non-trivial scalar field profile, and thus are different than in GR.

The Gauss-Bonnet term~\eqref{eq:gauss-bonnet-term} is quadratic in the curvature and thus is negligible in weak-field
regions of spacetime. Thus, EdGB gravity is not constrained by binary pulsar observations: the deviations from GR
only show up in very strong-field processes such as the merger and the ringdown of a binary BH coalescence. This makes
EdGB gravity a perfect candidate for gravitational spectroscopy.  Currently, the strongest bound on the coupling
constant $\alpha_{\rm GB}$ arises from the theoretical bound in Eqs.~\eqref{eq:zeta}, \eqref{eq:zetacond}: the existence
of the lightest BH observed, J1655-40, with mass $M\simeq5.4\,M_\odot$ implies~$\sqrt{\alpha_{\rm GB}}<6.6$ Km; if the
secondary object in GW190814, with mass $M\simeq2.6M_\odot$, turns out to be a BH, then~$\sqrt{\alpha_{\rm GB}}<3.3$
Km\,\cite{Carullo:2021dui}\,\footnote{For a comparison with other bounds in the literature arising from astrophysical
  observations~\cite{Yagi:2012gp,Seymour:2018bce}, taking into account differences in notations and conventions,
  see~\cite{Witek:2018dmd}.}.

QNMs of static, non-rotating BHs in EdGB gravity have been computed in~\cite{Salcedo2016}. The modes of rotating BHs
have only been estimated using the so-called ``geodesic
correspondence''~\cite{Ferrari:1984zz,Cardoso:2008bp,Yang:2012he}, which states that the QNMs can be approximately
computed in terms of the orbital frequencies near the light ring, and the approximation is better for larger values of
the harmonic index $l$. The geodesic correspondence has been formally proved for Kerr BHs in GR, but there is no proof
of it in modified gravity theories. Moreover, while the geodesic approximation predics the QNMs to be isospectral -
i.e., to coincide for perturbations with polar and axial parities - we know that in theories such as EdGB gravity and
Chern-Simons gravity~\cite{Alexander:2009tp} the QNMs are {\it not} isospectral~\cite{Molina:2010fb,Salcedo2016}, thus
the geodesic approximation is violated (see also~\cite{Moura:2021eln} for a computation of the QNMs of non-rotating BHs
in EdGB gravity within the geodesic approximation). In particular, it is reasonable to expect such violation when the
spacetime metric is coupled to additional degrees of freedom, e.g. to a scalar field, as for perturbations with polar
parity in EdGB gravity, and for perturbations with axial parity in Chern-Simons gravity.

We want to compute the QNMs of stationary, rotating BHs in EdGB gravity. To this aim, we perform a {\it slow-rotation
  expansion}, as in~\cite{Hartle2,Pani2012scalar,Pani:2013ija}. In order to test our approach, we first compute the QNMs
of a test scalar field on a BH, and compare our results with those of~\cite{Cano:2020cao}, where the dynamics of a test
scalar field on rapidly rotating BHs in EdGB gravity has been studied. We compute the QNMs of a scalar field up to order
$O({\bar a}^2)$, finding the same results as~\cite{Cano:2020cao} in the low-spin limit. We then compute gravitational
perturbations up to order $O(\bar a)$. 

We remark that most of the rotational corrections on the background metric appear at second order in the spin (the
metric at $O(\bar a)$ only acquires the frame-dragging $g_{t\varphi}$ component). Thus, there is no guarantee that the
main rotational corrections to the QNMs are captured by our $O(\bar a)$ analysis. Therefore, the results of this article
should be considered accurate for $\bar a\ll1$, while for larger values of the spin they are just an estimate; our
computation should be considered as a first step towards the computation of the QNMs of rotating BHs in EdGB gravity.

We use geometric units, such that $G=c=1$. In Section~\ref{sec:stationaryebhdgb} we describe static (non-rotating) and
stationary (rotating) BHs in EdGB gravity. In Section~\ref{sec:perturbations} we discuss perturbations of the stationary
BH background, order by order in $\bar a$. Finally, in Section~\ref{sec:concl} we draw our conclusions. In the
Appendices we describe the perturbation equations and their derivation, and discuss how we define the tortoise
coordinate for rotating BHs in EdGB gravity.

\section{Stationary black holes in  Einstein-dilaton Gauss-Bonnet gravity} \label{sec:stationaryebhdgb}
The equations of motion obtained from the action~\eqref{eq:S_edgb} in vacuum are
\begin{align}
  &\frac{1}{\sqrt{-g}}\partial_\mu (\sqrt{-g} g^{\mu \nu} \partial_\nu \phi) =
  \frac{\alpha_{\rm GB}}{4} e^\phi \mathcal{R}^2_{GB}
\label{eq:scalar}\\
&G_{\mu \nu}=\frac{1}{2} \partial_\mu \phi \partial_\nu \phi -
\frac{1}{4} g_{\mu\nu} (\partial_\rho \phi) (\partial^\rho \phi) - \alpha_{\rm GB} \mathcal{K}_{\mu\nu} \label{eq:metric}
\end{align}
where 
\begin{equation}
  \mathcal{K}_{\mu\nu}= \frac{1}{8}(g_{\mu\rho} g_{\nu\sigma} + g_{\mu\sigma} g_{\nu\rho})
  \epsilon^{\delta \sigma \gamma \alpha} \nabla_\beta \left({\tilde{R}^{\rho \beta}}_{ ~~\gamma \alpha}e^\phi \partial_\delta\phi\right)\,,
\label{eq:tensore_K}
\end{equation}
${\tilde{R}^{\mu \nu}}_{ ~~\rho \sigma} = \epsilon^{\mu\nu \delta \gamma}R_{\delta \gamma \rho \sigma}$, and
$\epsilon^{\mu\nu\delta\gamma}$ is the Levi-Civita tensor.

The solution describing a static, spherically symmetric BH in EdGB gravity has been derived in~\cite{Kanti1995} (see
also~\cite{PaniCardoso2009}). The spacetime metric of this solution can be written as:
\begin{equation}
ds^2=-A(r)dt^2+\frac{dr^2}{B(r)}+r^2d\Omega^2\label{eq:staticmetric}
\end{equation}
where $d\Omega^2=d\theta^2+\sin^2\theta d\varphi^2$. The functions $A(r)$, $B(r)$ and the scalar field $\phi(r)$ can be
found by numerical integration~\cite{Kanti1995,PaniCardoso2009}. By imposing asymptotic flatness and an asymptotically
vanishing scalar field, the solution is characterized by its ADM mass $M$ and its scalar charge $D$, which can be read
from the asymptotic behaviour of the metric and of the scalar field:
\begin{align}
  A&=1-\frac{2M}{r}+O\left(\frac{1}{r^3}\right)\nonumber\\
  \phi&=\frac{D}{r}+\frac{DM}{r^2}+O\left(\frac{1}{r^3}\right)\,.
\end{align}
The solution depends on the coupling constant $\alpha_{\rm GB}$ through the dimensionless coupling parameter
\begin{equation}
  \label{eq:zeta}
  \zeta=\frac{\alpha_{\rm GB}}{M^2}\,.
\end{equation}
Static black hole solutions exists for
\begin{equation}
  0\le\zeta\lesssim\zeta_{\rm max}\simeq 0.691\,;\label{eq:zetacond}
\end{equation}
for larger values of $\zeta$, it is impossible to impose regular boundary conditions to the field equations (the black
hole becomes a naked singularity~\cite{Sotiriou:2014pfa}). Note that the corresponding bound for rotating BHs is even
stronger~\cite{Kleihaus:2015aje}.

It is worth noting that the scalar charge $D$ is a so-called {\it secondary hair}, i.e. it is not an independent quantity:
\begin{equation}
  \frac{D}{M}=\frac{\zeta}{2}+\frac{73}{60}\zeta^2+O(\zeta^3)\,.
\end{equation}
Similarly, the horizon radius $r_{\rm h}$, defined by the condition $A(r_{\rm h})=B(r_{\rm h})=0$, is a function of the
mass and of the coupling $\zeta$: $r_{\rm h}=2M(1-49/1280\,\zeta^2)+O(\zeta^3)$.
Thus, for each value of $\zeta$ in the range~\eqref{eq:zetacond} and for each value of the mass there is one unique
static, spherically symmetric BH solution.

The field equations have also been solved perturbatively in the coupling parameter $\zeta$~\cite{Yunes:2011we} (see also, finding
an analytic expression for the metric and the scalar field, which can be written as:
\begin{align}
A(r)&=1-\frac{2M}{r}+\sum_{j=2}^{N_\zeta}\zeta^jA^{(j)}(r)\nonumber\\
B(r)&=1-\frac{2M}{r}+\sum_{j=2}^{N_\zeta}\zeta^jB^{(j)}(r)\nonumber\\
\phi(r)&=\sum_{j=1}^{N_\zeta}\zeta^j\phi^{(j)}(r)
\end{align}
where the functions $A^{(j)}(r)$, $B^{(j)}(r)$, $\phi^{(j)}(r)$ can be written as expansions in powers of $1/r$. The
bound~\eqref{eq:zetacond} guarantees that this expansion (possibly including higher-order terms in $\zeta$) is accurate.

The solution describing a stationary, rotating BH in EdGB gravity has been found
numerically~\cite{Kleihaus:2011tg,Kleihaus:2015aje}, by solving non-perturbatively the field equations, and
analytically~\cite{PaniCardoso2009,Maselli} in terms of a perturbative expansion in the coupling parameter and in the
spin ${\bar a}=J/M^2$, where $J$ is the angular momentum of the black hole:
\begin{align}
  ds^2 &= - A(r)[1+2h(r,\theta)] dt^2 + \frac{1}{B(r)}[1+2p(r,\theta)]dr^2 \nonumber\\
  &+ r^2 [1+2k(r,\theta)] \left[d\theta^2+\sin^2\theta(d\varphi-\varpi(r,\theta) dt)^2\right]\label{eq:metricslowrot}
\end{align}
with 
\begin{align}
  & \varpi = \sum_{j=2}^{N_\zeta}  \sum_{n=1,3,5...}^{N_{\bar a}} \sum_{l =1,3,5...}^{n}\zeta^j {\bar a}^n
  \omega_l ^{(nj)}(r) \left[ \frac{1}{\sin\theta}\frac{dP_l (\theta)}{d\theta}\right] \nonumber\\
  & h = \sum_{j=2}^{N_\zeta}\sum_{n=2,4,...}^{N_{\bar a}} \sum_{l =0,2,4...}^{n}\zeta^j
  {\bar a}^n h_l ^{(nj)}(r) P_l (\theta) \nonumber\\
  & p = \sum_{j=2}^{N_\zeta}\sum_{n=2,4,...}^{N_{\bar a}} \sum_{l =0,2,4...}^{n}\zeta^j
  {\bar a}^n p_l ^{(nj)}(r) P_l (\theta) \nonumber\\
	& k = \sum_{j=2}^{N_\zeta}\sum_{n=2,4,...}^{N_{\bar a}} \sum_{l =0,2,4...}^{n}\zeta^j {\bar a}^n k_l ^{(nj)}(r) P_l (\theta)
	\label{eq:expansionmetricslowrot}
\end{align}
and
\begin{equation}
\phi(r)=\sum_{j=1}^{N_\zeta}\sum_{n=0,2,4,...}^{N_{\bar a}} \sum_{l =0,2,4...}^{n}\zeta^j
  {\bar a}^n \phi_l ^{(nj)}(r) P_l (\theta)\label{eq:scalarslowrot}
\end{equation} 
where $P_l(\theta)$ are the Legendre polynomials, $N_\zeta$, $N_{\bar a}$ are the orders of the expansions in the
coupling and in the spin, respectively, and the functions $\omega_l ^{(nj)}(r)$, $h_l ^{(nj)}(r)$, $p_l ^{(nj)}(r)$,
$k_l^{(nj)}(r)$, $\phi_l ^{(nj)}(r)$ can be written as expansions in powers of $1/r$.  Their explicit expressions are
given in~\cite{Maselli} and, up to $n=2$ and $j=6$, in the supplemental {\sc Mathematica} noteboox~\cite{notebook}.
Similarly, the horizon radius $r_h$, the scalar charge $D$ and the maximum allowed coupling $\zeta_{\rm max}$ acquire
spin corrections with respect to the non-rotating case discussed in Sec.~\ref{sec:stationaryebhdgb}; they can be
expressed as expansions in $\zeta$ and ${\bar a}$ as well.

In the following we shall consider as a background solution the expansion~\eqref{eq:metricslowrot},
\eqref{eq:expansionmetricslowrot}, \eqref{eq:scalarslowrot} with $N_\zeta=6$ and $N_{\bar a}=2$~\cite{Maselli}. An
estimate based on the subsequent terms in the perturbation expansion shows that the truncation error on quantities
characterizing the background (such as the location of the horizon and the innermost stable circular orbit) is
$\lesssim2\%$ for $\bar a\le 0.7$ and $\zeta\le0.6$.
\section{Black hole perturbations}\label{sec:perturbations}
We shall consider perturbations of rotating BHs in EdGB gravity:
\begin{align}
  g_{\mu\nu}&=g^{(0)}_{\mu\nu}+h_{\mu\nu}\nonumber\\
  \phi&=\phi^{(0)}+\delta\phi\label{eq:defpert}
\end{align}
where $g^{(0)}_{\mu\nu}$, $\phi^{(0)}$ are given by Eqs.~\eqref{eq:metricslowrot}, \eqref{eq:expansionmetricslowrot},
\eqref{eq:scalarslowrot}.
\subsection{General equations}\label{sec:geneq}
We expand the perturbations of the metric tensor and of the scalar field in tensor spherical
harmonics. The scalar field perturbation is expanded as:
\begin{equation}
  \delta\phi(t,r,\theta,\varphi)=\frac{1}{r}\Phi^{lm}(r)Y^{lm}(\theta,\varphi)e^{-i\omega t}\,.\label{eq:exp_scal}
\end{equation}
The metric perturbations have contributions with polar and axial parities,
$h_{\mu\nu}=h^{\rm pol}_{\mu\nu}+h^{\rm ax}_{\mu\nu}$, and
\begin{align}
  &h^{\rm pol}_{\mu\nu}(t,r,\theta,\varphi)dx^\mu dx^\nu=\nonumber\\
  &\left[A(r)H_0^{lm}(r)dt^2+2H_1^{lm}(r)dtdr+B^{-1}(r)H_2^{lm}(r)dr^2\right.\nonumber\\
    &\left.+K^{lm}(r)(dr^2+\sin^2\theta d\varphi^2)\right]Y^{lm}(\theta,\varphi)e^{-i\omega t}\,,\label{eq:exp_pol}\\
&h^{\rm ax}_{\mu\nu}dx^\mu dx^\nu=2(h_0^{lm}(r)dt+h_1^{lm}(r)dr)\nonumber\\
  &\times(S_{\theta}(\theta,\varphi)d\theta+S_\varphi(\theta,\varphi)d\varphi)e^{-i\omega t}\label{eq:exp_ax}
\end{align}
  where $(S^{lm}_\theta,S^{lm}_\varphi)= (-(\sin\theta)^{-1}Y^{lm}_{,\varphi},\sin\theta Y^{lm}_{,\theta})$.
By replacing this expansion in the field equations~\eqref{eq:scalar}, \eqref{eq:metric} we find a set of partial
differential equations in $r$ and $\theta$ (see Appendix~\ref{app:eq}). Due to the symmetry of the background, the
dependence on $t$ and $\varphi$ factors out as $\sim e^{i(m\varphi-\omega t)}$: equations with different values of
$m,\omega$ are decoupled.

Following e.g.~\cite{Kojima1992} (see also~\cite{AdvancedMethods}), we perform a projection of these equations on tensor
spherical harmonics, finding a system of ordinary differential equations in $r$, up to second order in the spin. Since
the background is not spherically symmetric, equations with different values of $l$ are coupled; the general structure
of this system is:
\begin{align}
  &{\cal P}_{lm}+{\bar a}\,m\,\bar{\cal P}_{lm}+{\bar a}(Q_{lm}\tilde{\cal A}_{l-1\,m}+
  {\bar a}(Q_{l+1\,m}\tilde{\cal A}_{l+1\,m})\nonumber\\
&+{\bar a}^2(Q_{l-1\,m}Q_{lm}{\hat P}_{l-2\,m}+Q_{l+2\,m}Q_{l+1\,m}{\hat P}_{l+2\,m})+O({\bar a}^3)\nonumber\\
&=0&\nonumber\\
  &{\cal A}_{lm}+{\bar a}\,m\,\bar{\cal A}_{lm}+{\bar a}(Q_{lm}\tilde{\cal P}_{l-1\,m}+
  {\bar a}(Q_{l+1\,m}\tilde{\cal P}_{l+1\,m})\nonumber\\
&+{\bar a}^2(Q_{l-1\,m}Q_{lm}{\hat A}_{l-2\,m}+Q_{l+2\,m}Q_{l+1\,m}{\hat A}_{l+2\,m})+O({\bar a}^3)\nonumber\\
&=0\label{generalstructure}
\end{align}
where $Q_{lm}$ is given in Eq.~\eqref{eq:defQ} and ${\cal P}_{lm}$, $\bar{\cal P}_{lm}$, $\hat{\cal P}_{lm}$, (${\cal
  A}_{lm}$, $\bar{\cal A}_{lm}$, $\hat{\cal A}_{lm}$) are combinations of the polar perturbation functions $H_0^{lm}$,
$H_1^{lm}$, $H_2^{lm}$, $K^{lm}$, $\Phi^{lm}$ (of the axial perturbation functions $h_0^{lm}$, $h_1^{lm}$); see
Appendix~\ref{app:eq} and the {\sc Mathematica} notebook in the supplemental material~\cite{notebook}
for further details. The coefficients of these functions in the combinations ${\cal P}_{lm}$,
${\cal A}_{lm}$, etc. do not depend on the harmonic index $m$.

The structure of the equations~\eqref{generalstructure} critically depends on the order of the spin:
\begin{itemize}
\item At zero-th order in the spin, the equations reduce to ${\cal P}_{lm}=0$, ${\cal A}_{lm}=0$. Neglecting the spin in
  the background as well, these equations reduce to the perturbation equations of static, spherically symmetric BHs
  derived in~\cite{Salcedo2016}.
\item At first order in the spin, polar perturbations with $l$ are coupled to axial perturbations with $l\pm1$, and
  vice versa. Moreover, polar (axial) perturbations are coupled to perturbations having the same $l$ and the same
  parity; the latter couplings are proportional to $m$.
\item At second order of the spin, new terms appear coupling polar (axial) perturbations with $l$ and perturbations
  having the same parity and $l\pm2$.
\end{itemize}
\subsection{Quasi-normal modes}\label{sec:qnm}
The QNMs are free oscillations of the BH spacetime (see e.g.~\cite{Kokkotas:1999bd,Ferrari:2007dd,Berti:2009kk} and
references therein). They are solutions of the (scalar or gravitational) perturbation equations with Sommerfeld boundary
conditions (outgoing waves at infinity and ingoing waves at the horizon).

With an appropriate definition of the tortoise coordinate $r_*$ (see Appendix~\ref{app:tortoise}), all scalar
($\Phi^{lm}(r)$) and gravitational ($H^{lm}_1(r)/r$, $K^{lm}(r)$, etc.) perturbation functions behave at
the horizon and at infinity as
\begin{align}
&A_{\rm in}^{lm}e^{-ik_{\rm H}r_*}+A_{\rm out}^{lm}e^{ik_{\rm H}r_*}~\,~~~(r\to r_{\rm h})\nonumber\\
&A_{\rm in}^{lm}e^{-i\omega r_*}+A_{\rm out}^{lm}e^{i\omega r_*}~~~~~~~(r\to \infty)\,,\label{eq:bc0}
\end{align}
where 
\begin{equation}
  k_{\rm H}=\omega-m\Omega_{\rm H}~~~{\rm with}~~~\Omega_{\rm H}=-\lim_{r\to r_{\rm h}}\frac{g_{t\varphi}}{g_{\varphi\varphi}}\,.
\label{eq:defkh}
\end{equation}
We remark that the couplings between perturbations with different harmonic indices $l$, and the coupling between scalar
and gravitational perturbations, are subleading at the horizon and at infinity. Moreover, the equations for gravitational
perturbations, at the horizon and at infinity, can be combined into a second-order differential equation with the same
structure as the scalar field equation, i.e.
\begin{align}
  Z^{lm}_{,r_*r_*}+k_H^2Z^{lm}&=O(r-r_{\rm h})~~~~~(r\to r_{\rm h})\nonumber\\
   Z^{lm}_{,r_*r_*}+\omega^2Z^{lm}&=O\left(\frac{1}{r^2}\right)~~~~~(r\to\infty)\,.\label{eq:expwave}
\end{align}
where $Z^{lm}$ is either $\Phi^{lm}$ or $K^{lm}$.

The Sommerfeld boundary conditions are then
\begin{align}
  Z^{lm}&\sim e^{-ik_{\rm H}r_*}~\,~~~{\rm at}~~(r\to r_{\rm h})\nonumber\\
  Z^{lm}&\sim e^{\,i\omega r_*}~~~~~~~{\rm at}~~(r\to \infty)\,.\label{eq:bc}
\end{align}
They can be satisfied for a discrete set of complex frequencies
\begin{equation}
\omega=\omega_R+i\,\omega_I\,:
\end{equation}
the QNMs of the BH.
\subsection{Gravitational perturbations at zeroth-order in the spin}\label{sec:spherical}
To begin with, we compute the QNMs of a non-rotating BH in EdGB gravity. We reproduce the results of~\cite{Salcedo2016},
where these modes were first computed.

We only consider perturbations with polar parity, i.e.  $H_0$, $H_1$, $H_2$, $K$, $\Phi$; we leave implicit the dependence on the
harmonic indices $l,m$. The perturbation equations give $H_0$, $H_2$ as algebraic expressions in terms of $K$,
$H_1$. The remaining equations can be written as a linear, first-order ODE system for $H_1$, $K$, the scalar field
$\Phi$ and its first derivative $\xi\equiv\Phi'$:
\begin{equation}
\frac{d}{dr} \bm{\Psi} + \bm{V}\bm{\Psi} = \bm{0}
\label{eq:zerilli_vett_om}
\end{equation}
were 
\begin{equation}
\bm \Psi = 
\begin{pmatrix}
H_1\\
K\\
\Phi\\
\xi\\
\end{pmatrix}
\label{eq:psi_edgb}
\end{equation}
and $\bm{V}$ is a four-dimensional square matrix, whose components $V_{ij}(r)$ $(i,j=1,\dots4)$ are linear in
$H_1$, $K$, $\Phi$, $\xi$. Their expansions in the coupling parameter $\zeta$ up to $O(\zeta^6)$ are given in the supplemental
{\sc Mathematica} notebook~\cite{notebook}.  At the horizon and at infinity it reduces to a diagonal form of the
kind~\eqref{eq:expwave}, where the tortoise coordinate is given by (see Appendix~\ref{app:tortoise})
\begin{equation}
\frac{dr}{dr_*}=\sqrt{A(r)B(r)}\,,	\label{eq:tortoise-static}
\end{equation}
and (since the BH does not rotate) $k_{\rm H}=\omega$.

The space of solutions of Eq.~(\ref{eq:zerilli_vett_om}) is a four-dimensional vector space. We can find two independent
solutions satisfying ingoing boundary conditions ($\sim e^{-i\omega r_*}$) at the horizon; we choose two of them,
denoted as $\bm\Psi^-_a$, $\bm\Psi^-_b$. Similarly, we can find two independent solutions satisfying outgoing boundary
conditions ($\sim e^{i\omega r_*}$) at infinity; we choose two of them, denoted as $\bm\Psi^+_a$, $\bm\Psi^+_b$. We find
these solutions by numerical integration of $\bm\Psi^+_i$ ($i=a,b$) from infinity to a matching point $r=r_m$, and by
numerical integration of $\bm\Psi^-_i$ from the horizon to $r=r_m$; the results do not depend on the choice of $r_m$.

If $\omega$ is the frequency of a QNM, these solutions are linearly dependent, and thus the matrix
\begin{equation}
\bm X=
\begin{pmatrix}
H_{1a}^- & H_{1b}^- & H_{1a}^+ & H_{1b}^+\\
K_a^- & K_b^- & K_a^+ & K_b^+\\
\Phi_{a}^- & \Phi_{b}^- & \Phi_{a}^+ & \Phi_{b}^+\\
\xi_a^- & \xi_b^- & \xi_a^+ & \xi_b^+\\
\end{pmatrix}
\label{eq:X_edgb}
\end{equation}
is degenerate. Thus, the QNMs are the solutions of the equation
\begin{equation}
{\rm det}\bm X(\omega^{nl})=0\,.
\label{eq:detX}
\end{equation}
Our results agree with the fits  obtained in~\cite{Salcedo2016}
\begin{align}
& \frac{\omega^{nl}_R}{\omega^{nl}_R(\zeta=0)} = 1 + \sum_{j=1}^4 R^{nl}_j \zeta^j \\
& \frac{\omega^{nl}_I}{\omega^{nl}_I(\zeta=0)} = 1 + \sum_{j=1}^4 I^{nl}_j \zeta^j 
\label{eq:Sal-fit}
\end{align}
up to $1\%$ for $\zeta\le0.5$. Note that the expansion in~\cite{Salcedo2016} include terms up to $O(\zeta^4)$;
therefore, in our comparison we have included terms up to $O(\zeta^4)$.
\subsection{Test scalar field at second order in the spin}\label{sec:scalar2}
As an additional test of our approach, we computed the QNMs of a test scalar field on a rotating BH background in EdGB
gravity, up to second order in the rotation, comparing our results with those of~\cite{Cano:2020cao}. We remark that
while in this paper we consider scalar and gravitational perturbations of slowly rotating BHs in terms of a perturbative
expansion in the spin ${\bar a}$, Ref.~\cite{Cano:2020cao} considers a test scalar field on a rapidly rotating BH
background, through an extension of the Teukolsky formalism to EdGB gravity. Thus, we can compare our results with those
of~\cite{Cano:2020cao} for a scalar field only, and for small values of the spin.

We consider a massless scalar field $\phi$ satisfying the Klein-Gordon equation
\begin{equation}
  \nabla_\mu\nabla^\mu\phi=0\label{eq:KG}
\end{equation}
in the background~\eqref{eq:expansionmetricslowrot}, up to second order in the rotation rate. By replacing the
background metric and the expansion of the scalar field in spherical harmonics:
$\phi(t,r,\theta,\varphi)=\frac{1}{r}\phi^{lm}(t,r)Y^{lm}(\theta,\varphi)$ in Eq.~\eqref{eq:KG}, looking for solutions
with the form $\phi^{lm}(t,r)=\phi^{lm}(r)e^{-i\omega t}$, and projecting on the basis of spherical harmonics, we find a set of
coupled equations with the form (we leave implicit the index $m$)
\begin{equation}
\frac{d}{dr} \bm{\Psi}^{l} + \bm{V}_{l}\bm{\Psi}^{l} = \bm{S}_{l-2} \bm{\Psi}^{l-2}+\bm{S}_{l+2}\bm{\Psi}^{l+2}
\label{eq:sfKG_vett}
\end{equation}
where $l=0,1,\dots$, 
\begin{equation}
\bm \Psi^{l} = 
\begin{pmatrix}
\phi^{l}\\
{\phi}'^{l}\\
\end{pmatrix}\,,
\label{eq:psi_edgb_KG}
\end{equation}
the prime denotes differentiation with respect to $r$, and $\bm{V}_l= \bm{V}_l^{(0)}+ {\bar a} m \bm{V}_l^{(1)}$,
$\bm{S}_l = {\bar a}^2 \bm{S}^{(2)}_l $ are two-dimensional square matrices. The expansions in the coupling parameter
$\zeta$ up to $O(\zeta^4)$ of the components of these matrices are given in the supplemental {\sc Mathematica}
notebook~\cite{notebook}.

We truncate the harmonic expansion to $l_{max}=4$, and solve the three coupled equations~(\ref{eq:sfKG_vett}) with
$l=0,2,4$ (the equations with odd values of $l$ are decoupled from those with even values, and correspond to different
modes). We can further recast this system as
\begin{equation}
\frac{d}{dr} \bm{Z} + \bm{W}\bm{Z} = 0
\label{eq:sfKG_vett_2}
\end{equation}
where $\bm Z$ is the six-dimensional vector
\begin{equation}
\bm Z = 
\begin{pmatrix}
\bm{\Psi}^0\\
\bm{\Psi}^2\\
\bm{\Psi}^4\\
\end{pmatrix}
\label{eq:Zeta_KG}
\end{equation}
and
\begin{equation}
\bm W=
\begin{pmatrix}
-\bm{V}_0 & \bm{S}_2 & \bm 0 \\
\bm{S}_0 & -\bm{V}_2 & \bm{S}_4 \\
\bm 0 & \bm{S}_2 & -\bm{V}_4\\
\end{pmatrix}\,.
\label{eq:W_KG}
\end{equation}
The quasi-normal modes satisfy ingoing boundary conditions at the horizon ($r_*\to-\infty$), $\bm\Phi\sim e^{-i k_H r_*}$
with $k_H$ given in Eq.~\eqref{eq:defkh}, and outgoing boundary conditions at infinity ($r_*\to\infty$),
$\bm\Psi\sim e^{i \omega r_*}$ (see Appendix~\ref{app:tortoise} for the definition of the tortoise coordinate).

Proceeding as in Sec.~\ref{sec:spherical}, we find (by direct integration from the horizon) three independent solutions
$\bm Z^{-}_i$ ($i=a,b,c$) satisfying the boundary conditions at the horizon, and (by direct integration from infinity)
three independent solutions $\bm Z^{+}_i$ satisfying the boundary conditions at infinity. Thus, we define the
six-dimensional square matrix
\begin{equation}
\bm X=
\begin{pmatrix}
\bm{\Psi}_a^{0-} & \bm{\Psi}_b^{0-} & \bm{\Psi}_c^{0-} & \bm{\Psi}_a^{0+} &\bm{\Psi}_b^{0+} &\bm{\Psi}_c^{0+} \\
\bm{\Psi}_a^{2-} & \bm{\Psi}_b^{2-} & \bm{\Psi}_c^{2-} & \bm{\Psi}_a^{2+} &\bm{\Psi}_b^{2+} &\bm{\Psi}_c^{2+} \\
\bm{\Psi}_a^{4-} & \bm{\Psi}_b^{4-} & \bm{\Psi}_c^{4-} & \bm{\Psi}_a^{4+} &\bm{\Psi}_b^{3+} &\bm{\Psi}_c^{4+} \\
\end{pmatrix}
\label{eq:X_KG}
\end{equation}
The QNMs are given by:
\begin{equation}
{\rm det}\bm X(\omega^{nlm})=0\,.
\label{eq:detX2}
\end{equation}
In order to estimate the contribution of the $O({\bar a}^2)$ terms, we repeat the computation by neglecting the quadratic
terms in the spin; in this case the right-hand side of Eq.~\eqref{eq:sfKG_vett} vanishes.

We then compare our results with those of~\cite{Cano:2020cao} in the limit of small spins.  Since in~\cite{Cano:2020cao} the
EdGB corrections are included up to $O(\zeta^2)$, we also compute the scalar field QNMs up to $N_\zeta=2$.  We define
the shifts
\begin{equation}
  f^{A}_{R,I}(\zeta,{\bar a}) = \lim_{\zeta\to0}\frac{\omega_{R,I}(\zeta,{\bar a}) -
    \omega_{R,I}(\zeta=0,{\bar a})}{\zeta^2\,\omega_{R,I}(\zeta=0,{\bar a})}
\label{eq:delta_comp_cano}
\end{equation}
describing the leading-order ($O(\zeta^2)$) EdGB corrections to the real and imaginary parts of the QNMs; $A=1$ ($A=2$)
refers to the computation of the QNMs up to $O(\bar a)$ (up to $O({\bar a}^2)$). We denote with $f_{R,I}^{\rm C}$ the
corresponding shifts obtained by the numerical fits in\,\cite{Cano:2020cao}.  In Table \ref{tab:comp_Cano} we show that
our results are in good agreement with those of~\cite{Cano:2020cao} for ${\bar a}\le0.1$ if the quadratic terms in the
spin are included; the agreement is much worst with the computation to $O(\bar a)$\,\footnote{This is not true for the
  smallest value of the spin in Table~\ref{tab:comp_Cano}, $\bar a=0.01$. We think this is due to the fact --~also
  remarked by the authors of\,\cite{Cano:2020cao}~-- that their fit is optimized to describe the entire range $0\le\bar
  a\le0.7$, and thus it may not coincide with a perturbative expansion for $\bar a\ll1$.}  , in particular for the
imaginary part.

\begin{table}[h]
  \caption{\textit{Leading-order EdGB corrections to the $n=0$, $l=m=2$ QNM, defined in Eq.~\eqref{eq:delta_comp_cano}, 
      up to first and to second order in the spin, and as computed in~\cite{Cano:2020cao}.}}
  \centering
	\begin{tabular}{ c | c c  c }
		\toprule
		${\bar a}$ & $f^{\rm C}_R$ & $f^1_R$   & $f_R^2$  \\
		\midrule
                0.01 & 0.013653 & 0.013667 & 0.013660 \\
                0.05 & 0.014273 & 0.014576 & 0.014274 \\
                0.1 &  0.015065 & 0.016565 & 0.015004\\
                \bottomrule
                \toprule
		${\bar a}$ & $f^{\rm C}_I$ & $f^1_I$   & $f_I^2$  \\		
		\midrule
                0.01 & -0.005178  &-0.005092 &  -0.005035\\
                0.05 & -0.005287 & -0.006337  & -0.005089\\
                0.1 &  -0.005410 & -0.007087 &  -0.005272 \\
		\bottomrule
	\end{tabular}
	\label{tab:comp_Cano}
\end{table}

\subsection{Gravitational perturbations at first order  in the spin}\label{sec:firstorder}
Finally, we studied QNMs of gravitational perturbations in EdGB gravity. Due to the complexity of the problem, this
computation has been carried out up to order $O({\bar a})$.
\subsubsection*{Computation of the quasi-normal modes}
We decompose the metric perturbations with polar and axial parities as in Eqs.~\eqref{eq:exp_pol}, \eqref{eq:exp_ax}, in
terms of the perturbation functions of polar parity $\{H_0^{lm}(t,r),H_1^{lm}(t,r),H_2^{lm}(t,r),K^{lm}(t,r)\}$ and of
axial parity $\{h_0^{lm}(t,r),h_1^{lm}(t,r)\}$.  Similarly, we expand the scalar field perturbation as in
Eq.~\eqref{eq:exp_scal}, in terms of the perturbation function (with polar parity) $\Phi^{lm}(t,r)$.

By replacing these expansions in the field equations, we find - at first order in the spin - a set of equations whose
general structure~\eqref{generalstructure} reduces to
\begin{align}
  &\mathcal{A}_{lm}+ {\bar a} m \bar{\mathcal{A}}_{lm}+ {\bar a} \left(Q_{lm} \tilde{\mathcal{P}}_{l-1\,m}
  +Q_{l+1\,m}\tilde{\mathcal{P}}_{l+1\,m}\right)=0 \label{eq:eq-gen-axial-1st}\\
  & \mathcal{P}_{lm} + {\bar a}  m \bar{\mathcal{P}}_{lm} + {\bar a} \left(Q_{lm}
  \tilde{\mathcal{A}}_{l-1\,m}+Q_{l+1\,m}\tilde{\mathcal{A}}_{l+1\,m}\right)=0\,. \label{eq:eq-gen-polar-1st}
\end{align}
In order to find the QNMs, we look for solutions of the perturbation equations of the form
$Z^{lm}(t,r)=Z^{lm}(r)e^{-i\omega t}$ (where $Z^{lm}$ is any of the scalar and gravitational perturbation functions).

As discussed in detail in~\cite{1993PThPh..90..977K,1993ApJ...414..247K,Pani2012scalar}, the couplings to the $l\pm1$
terms do not contribute to the QNM spectrum to first order in the spin. Indeed, due to the symmetry properties of
Eqns.~\eqref{eq:eq-gen-axial-1st}, \eqref{eq:eq-gen-polar-1st}, the QNM frequencies can be expanded as
\begin{align}
\omega=\omega_0+ {\bar a} \, m \omega_1+ {\bar a}^2 \,\omega_2 + O({\bar a}^3)\,.
\label{eq:qnms-expansion}
\end{align}
The coefficients of the perturbation functions in $\mathcal{A}_{lm}$, $\mathcal{P}_{lm}$, $\bar{\mathcal{A}}_{lm}$,
$\bar{\mathcal{P}}_{lm}$, $\tilde{\mathcal{A}}_{l\pm1\,m}$, $\tilde{\mathcal{P}}_{l\pm1\,m}$ do not depend on the
harmonic index $m$. Moreover, the terms $\bar{\mathcal{A}}_{lm}$, $\bar{\mathcal{P}}_{lm}$,
$\tilde{\mathcal{A}}_{l\pm1\,m}$, $\tilde{\mathcal{P}}_{l\pm1\,m}$ only depend on perturbations of order $O({\bar
  a}^0)$.  Thus, the QNM correction ${\bar a}m\omega_1$ is given by the terms $ {\bar a} m \,\bar{\mathcal{A}}_{lm}$, $
{\bar a} m \,\bar{\mathcal{P}}_{lm}$ in the equations, while it is not affected by the terms
$\tilde{\mathcal{A}}_{l\pm1\,m}$, $\tilde{\mathcal{P}}_{l\pm1\,m}$.

For this reason, we shall neglect the $\tilde{\mathcal{A}}_{l\pm1\,m}$, $\tilde{\mathcal{P}}_{l\pm1\,m}$ terms in the
following; the general structure of the equations, then, reduces to:
\begin{align}
&\mathcal{A}_{lm}+ {\bar a} m \bar{\mathcal{A}}_{lm}=0 \label{eq:eq-gen-axial-1st-simp}\\
& \mathcal{P}_{lm} + {\bar a} m \bar{\mathcal{P}}_{lm}=0 \,.\label{eq:eq-gen-polar-1st-simp}
\end{align}
Following the notation of~\cite{Kojima1992}, and leaving implicit the harmonic indices $l,m$, the perturbation equations can be
written in the form (see Appendix~\ref{app:eq})
\begin{flalign}
&  A^{(I)} + i m C^{(I)}=0\nonumber\\
  &  l(l+1)\alpha^{(J)}- i m \left[\tilde{\beta}^{(J)} +
    {\hat \zeta}^{(J)}-(l-1)(l+2){\hat \xi}^{(J)}\right]=0 \nonumber\\
  &  l(l+1)\beta^{(J)} + i m \left[\tilde{\alpha}^{(J)} +
    {\hat \eta}^{(J)} + (l-1)(l+2){\hat \gamma}^{(J)}\right]=0\nonumber\\
&  l(l-1)(l+1)(l+2) \hat s - i m (l-1)(l+2)\hat f =0\nonumber\\
&  l(l-1)(l+1)(l+2) \hat t + i m (l-1)(l+2)\hat g  =0
\label{eq:1st-simp}
\end{flalign}
where $A^{(I)}$, $C^{(I)}$, $\alpha^{(J)}$, $\beta^{(J)}$, etc. ($I=0,\dots,4$, $J=0,1$) are combinations of the perturbation
functions and their derivatives, whose expansions in the coupling parameter $\zeta$ up to $O(\zeta^6)$ are given in 
the supplemental {\sc Mathematica} notebook~\cite{notebook}.

With appropriate combinations of the perturbation equations, we can find $H_0$ and $H_2$ as algebraic expressions in
terms of $H_1$ and $K$. Thus, calling $\xi=\Phi'$ and defining the vector quantity
\begin{equation}
\bm \Psi = 
\begin{pmatrix}
H_1\\
K\\
\Phi\\
\xi\\
\end{pmatrix}\,,
\label{eq:psi_edgb-2}
\end{equation}
we can cast our equations as
\begin{equation}
\frac{d}{dr} \bm{\Psi} + \bm{\hat{V}}\bm{\Psi} + {\bar a} \ m \  \bm{\hat{U}}\bm{\Psi} = \bm{0}\,,
\label{eq:vect-system-1st}
\end{equation}
where $\bm{\hat{V}}$ and $\bm{\hat{U}}$ are four-dimensional square matrices. The expansions in the coupling parameter
$\zeta$ up to $O(\zeta^6)$ of the components of these matrices are given in the supplemental {\sc Mathematica}
notebook~\cite{notebook}.  With an appropriate definition of the tortoise coordinate $r_*$ (see
Appendix~\ref{app:tortoise}), the perturbation functions behave at the horizon and an infinity as in Eq.~\eqref{eq:bc0}.
The QNMs satisfy ingoing boundary conditions at the horizon ($\sim e^{-i k_H r_*}$) with $k_H$ given in
Eq.~\eqref{eq:defkh}, and outgoing boundary conditions at infinity ($\sim e^{i \omega r_*}$).

We define a four-dimensional square matrix whose columns are two independent solutions satisfying the QNM boundary
conditions at the horizon (superscript $^{(-)}$), and two independent solutions satifying the boundary conditions at
infinity (superscript $^{(+)}$), evaluated at a matching point $r_m$:
\begin{equation}
\bm{X}=
\begin{pmatrix}
H_{1a}^- & H_{1b}^- & H_{1a}^+ & H_{1b}^+\\
K_a^- & K_b^- & K_a^+ & K_b^+\\
\Phi_{a}^- & \Phi_{b}^- & \Phi_{a}^+ & \Phi_{b}^+\\
\xi_a^- & \xi_b^- & \xi_a^+ & \xi_b^+\\
\end{pmatrix}\,.
\label{eq:X_edgb_2}
\end{equation}
The QNMs are found by imposing the condition
\begin{equation}
{\rm det} \bm{X}(\omega^{nlm})=0\,.
\label{eq:detX2bis}
\end{equation}
The QNM frequencies at first order in the rotation (see Eq.~(\ref{eq:qnms-expansion})) can be written as
\begin{equation}
\omega^{nlm}({\bar a},\zeta)=\omega^{nl}_0 (\zeta)+ {\bar a}\,m \omega^{nl}_1 (\zeta) +\mathcal{O}({\bar a}^2)
\label{eq:qnms-expansion-2}
\end{equation}
where $\omega_0$ is the QNM frequency in the static case. We determine the rotational corrections
$\omega_1^{nl}(\zeta)=\omega^{nl}_{1\,R}(\zeta)+i\,\omega^{nl}_{1\,I}(\zeta)$ by studying the $\bar a\to0$ limit of the
QNMs.

As discussed in~\cite{Salcedo2016} (see also~\cite{Cardoso:2009pk}), in modified gravity theories with a scalar field
coupled to the metric perturbations two classes of gravitational QNMs exist: the {\it gravitational-led} modes and the
{\it scalar-led} modes, whose frequencies tend, in the $\zeta\to0$ limit, to those of gravitational and scalar QNMs in
GR, respectively. However, in a realistic physical scenario the gravitational-led modes are expected to be excited with
much larger amplitudes than the scalar-led modes~\cite{Salcedo2016}. Therefore, we expect that only gravitational-led QNMs
are relevant for gravitational spectroscopy, and thus we shall only study this class of modes. For gravitational-led
modes we find that, at $\zeta\to0$, $\omega_1^{nl}$ give the rotational corrections of QNMs in Kerr spacetime,
e.g. ${\omega}^{02}_{1\,R}(\zeta=0)=0.0629$~\cite{bertiweb}.

\begin{figure}[th]
\includegraphics[width=8cm,trim=3 3 3 3,clip]{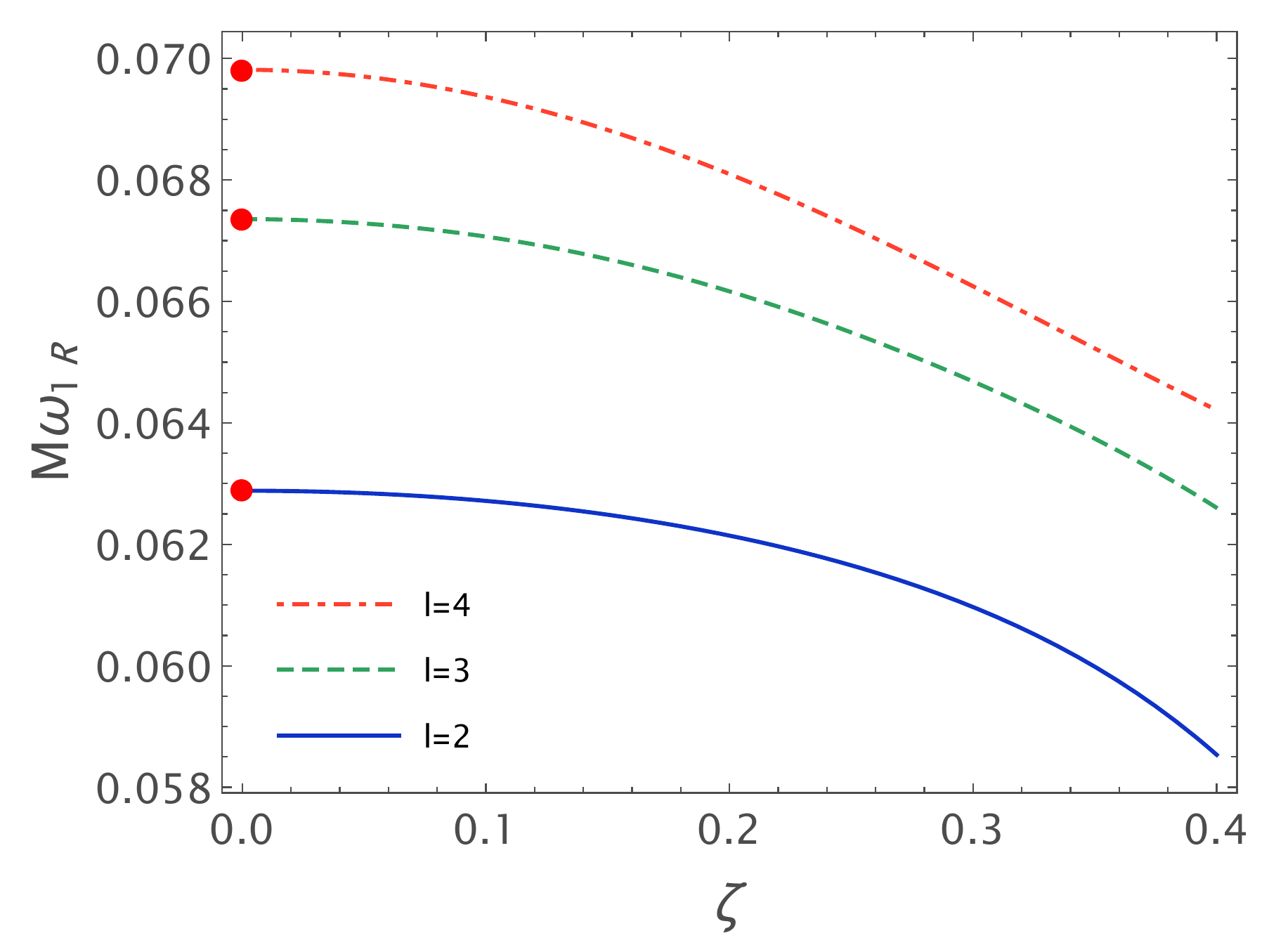}\vskip .2cm
\includegraphics[width=8cm,trim=3 3 3 3,clip]{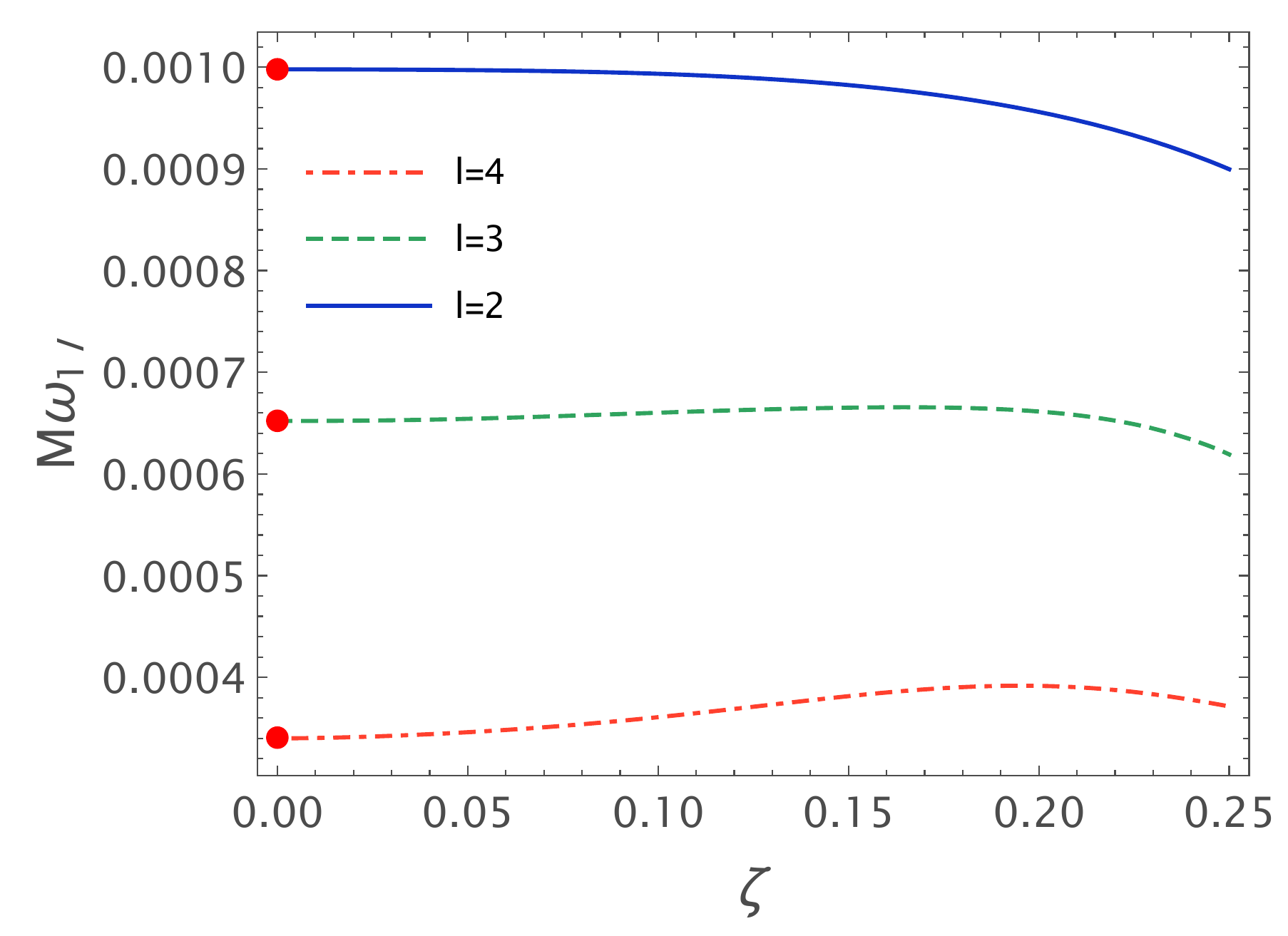}
\caption{First-order rotational corrections $\omega_1$ (see Eq.~\eqref{eq:qnms-expansion-2}) of the gravitational
  QNMs with $n=0$ and $l=2,3,4$, as functions of the coupling parameter $\zeta$. The small circles are the values
  corresponing to Kerr BHs. The real parts are shown in the upper panel, the imaginary parts in the lower panel.}
\label{fig:omega1}
\end{figure}

In Fig.~\ref{fig:omega1} we show the real and imaginary parts of $\omega_1$ for the fundamental ($n=0$) modes with
$l=2,3,4$, as functions of the dimensionless coupling parameter $\zeta$\,\footnote{Our numerical integration has
  convergence issues for $\omega_{1\,R}$ at $\zeta\gtrsim0.4$, and for $\omega_{1\,I}$ at $\zeta\gtrsim0.25$. Therefore,
  in Fig.~\ref{fig:omega1} we show the real part of the rotational correction for $\zeta\in[0,0.4]$, and the imaginary
  part for $\zeta\in[0,0.25]$.}. The curves for $n=0$, $l=2,3$ correspond to the following analytical fits:
\begin{align}
  M\omega^{02}_{1\,R}&=0.0629 - 0.0156 \zeta^2 - 0.00758 \zeta^3\nonumber\\
  &- 0.0644 \zeta^4 + 0.268 \zeta^5 - 0.603\zeta^6\nonumber\\  
  M\omega^{02}_{1\,I}&=0.00099 - 0.00110 \zeta^2 + 0.01864 \zeta^3 \nonumber\\
  &- 0.17271 \zeta^4 + 0.56422\zeta^5 - 0.8119 \zeta^6 \label{eq:fit}\\
M\omega^{03}_{1\,R}&=0.0674 - 0.0291 \zeta^2 + 0.0251 \zeta^3\nonumber\\
&- 0.3209 \zeta^4 + 1.1703 \zeta^5 - 1.3341 \zeta^6\nonumber\\  
M\omega^{03}_{1\,I}&=0.00065 + 0.00023 \zeta^2 + 0.0233 \zeta^3 \nonumber\\
&- 0.2832 \zeta^4 + 1.323 \zeta^5 - 2.442  \zeta^6\,.
\label{eq:fitl3}
\end{align}

\subsubsection*{Estimate of the truncation errors}
\begin{figure}[th]
\includegraphics[width=8cm,trim=3 3 3 3,clip]{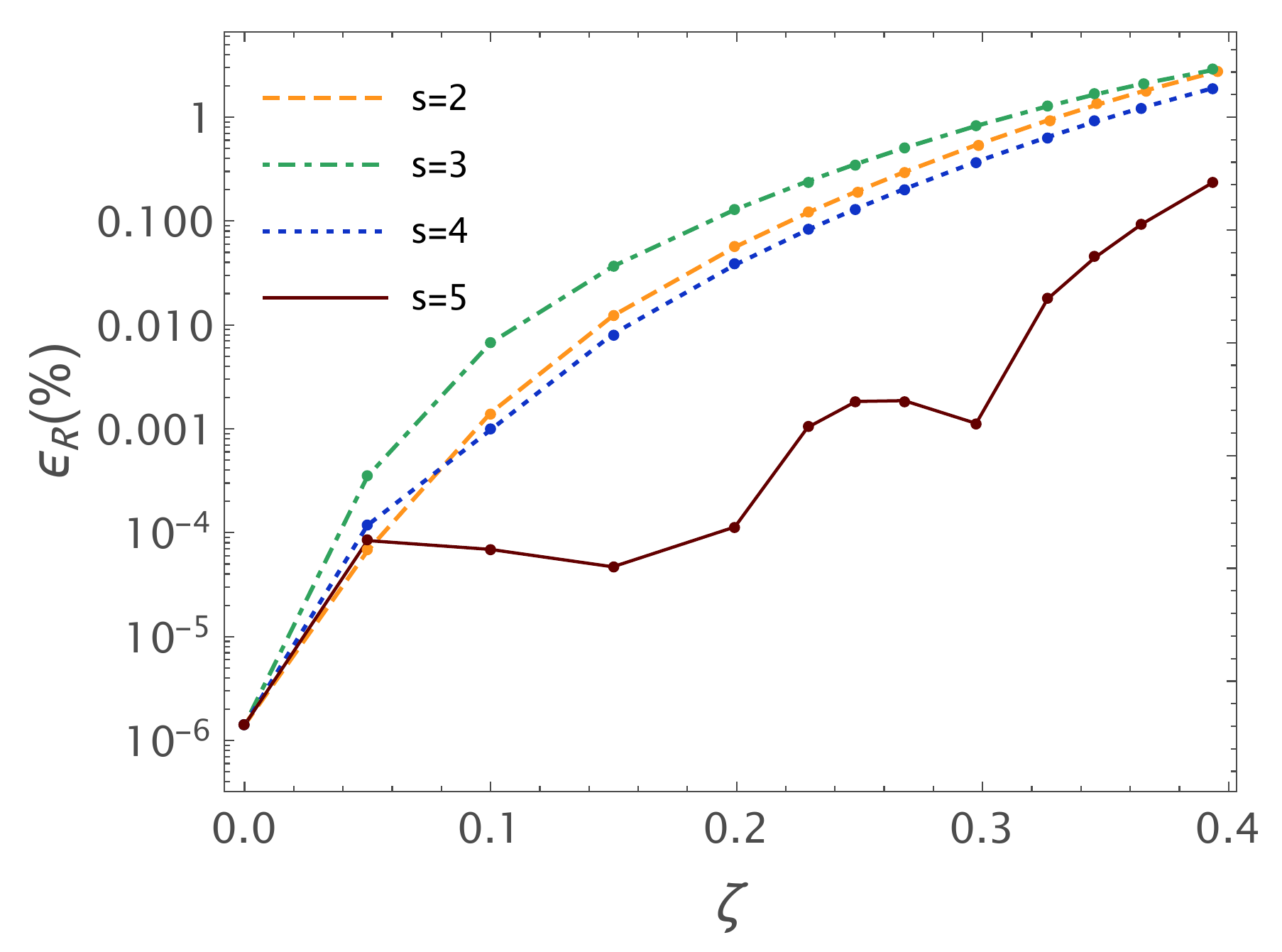}
\includegraphics[width=8cm,trim=3 3 3 3,clip]{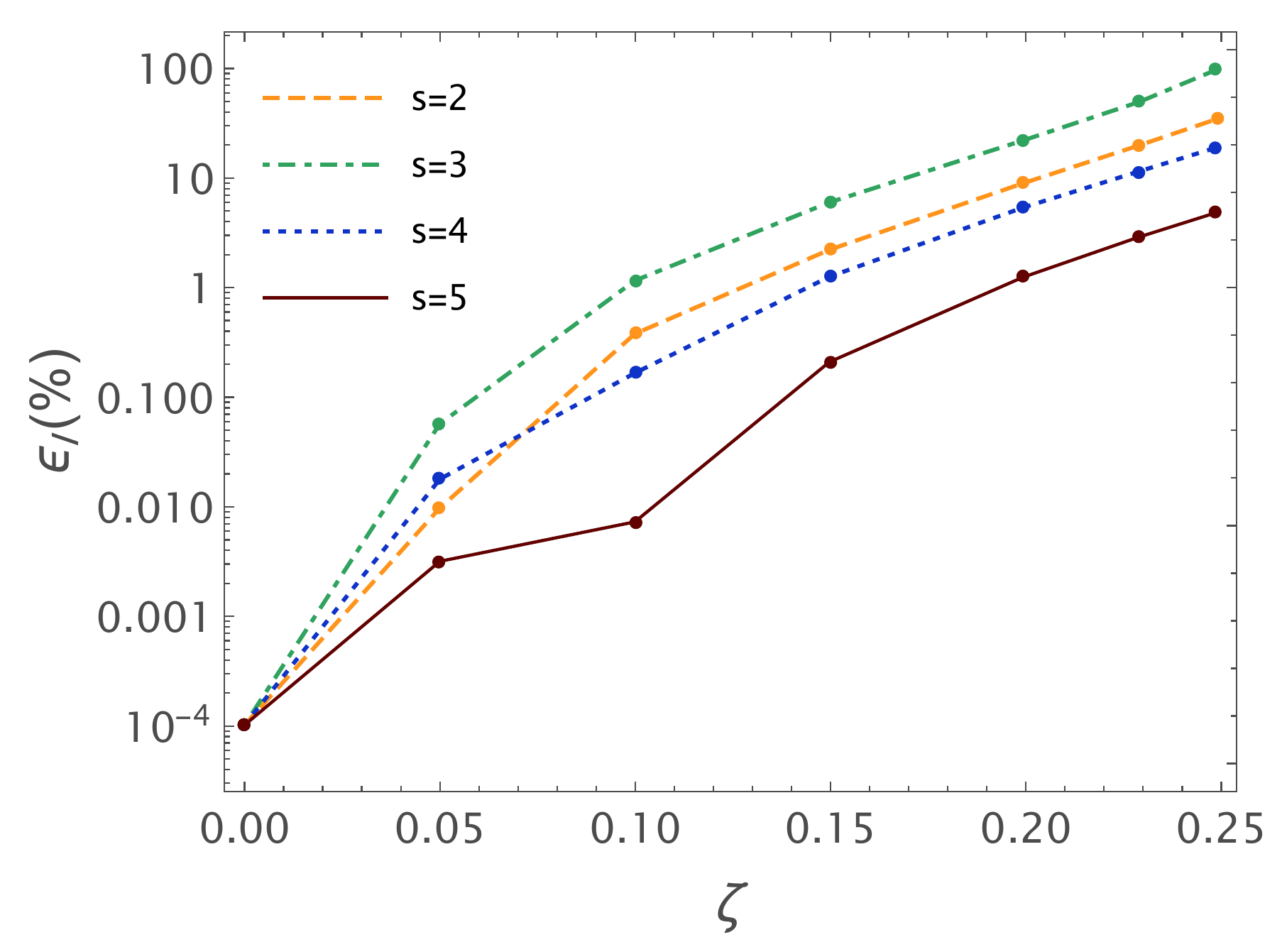}
\caption{Relative discrepancy (Eq.~\eqref{eq:discr}) between the estimate of $\omega_1$ for the $l=m=2$ fundamental mode
  computed to $O(\zeta^s)$ and to $O(\zeta^{s+1})$, for different values of the coupling parameter $\zeta$.}
\label{fig:discrepancy}
\end{figure}
In order to estimate the error in $\omega_1$ due to the truncation of the expansion in $\zeta$, we compare $\omega_1$
computed at $O(\zeta^s)$ and at $O(\zeta^{s+1})$. In Fig.~\ref{fig:discrepancy} we show the relative error
\begin{equation}
\epsilon^s_{R/I}=\frac{  |\omega_{1\,R/I}^{(s+1)}-  \omega_{1\,R/I}^{(s)}|}{  \omega_{1\,R/I}^{(s)}}\label{eq:discr}
\end{equation}
(where the superscript $(s)$ denotes the order of $\zeta$ included in the perturbative expansion),
for $s\le5$ and $\zeta\in[0,0.4]$ for the real part, $\zeta\in[0,0.25]$ for the imaginary part. This provides evidence
that the truncation error due to the expansion in $\zeta$ is always $\lesssim1\%$.

Conversely, it is impossible to provide a reliable estimate of the error due to the truncation at first order in the
spin. Yet, the integration of the test scalar field (see Table~\ref{tab:comp_Cano}) gives the indication that a
first-order computation may provide the leading-order contribution of the modes (in particular of the real
part). Generally speaking, corrections at order $O({\bar a}^2)$ may be qualitatively different from those at order
$O(\bar a)$; there is no guarantee, then, that the latter dominate the QNM frequencies unless the spin is very
low. Thus, our results should be considered as a first estimate of the rotational corrections to the QNMs in EdGB
gravity.
\subsubsection*{Discussion of the results}
\begin{figure}[th]
\includegraphics[width=8cm,trim=3 3 3 3,clip]{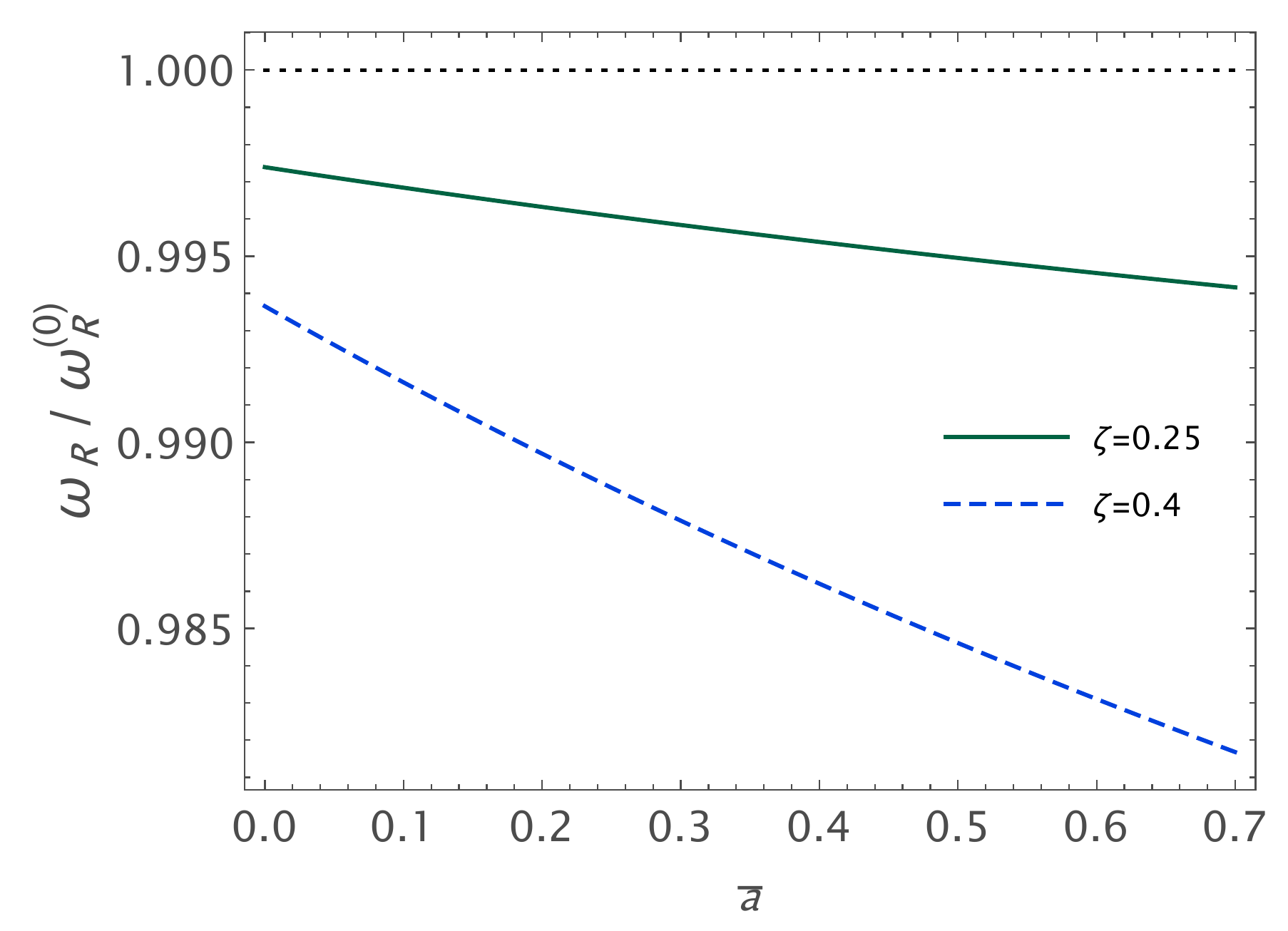}
\caption{Ratio between the (real part of the) $n=0$, $l=m=2$ QNM frequency, computed up to $O(\bar a)$, and the
  corresponding value in general relativity, as a function of the BH spin, for $\zeta=0.25$ and for $\zeta=0.4$.}
\label{fig:ratio}
\end{figure}

In order to assess how rotation affects the EdGB corrections to the QNMs, in Figure~\ref{fig:ratio} we show the ratio
between the QNM frequency $\omega^{nlm}$ (for $n=0$, $l=m=2$) in EdGB gravity and in general relativity, for different
values of $\zeta$. We only show the real part of the frequency, which we model as in Eq.~\eqref{eq:qnms-expansion-2}.

We consider values of the spin $\bar a\in[0,0.7]$ (the latter corresponding to the typical outcome of a binary BH
merger).  As discussed above, since we neglect terms $O({\bar a}^2)$, this model is accurate for $\bar a\ll1$, while it
should be only considered as an order-of-magnitude estimate of the corrections for $\bar a\lesssim0.7$.

As shown in Fig.~\eqref{fig:ratio}, rotation significantly magnifies the correction due to modified gravity. For
$\zeta=0.4$, the mode is shifted of $\sim4\%$ in a non-rotating BH, while the shift increases to $\sim18\%$ for a BH
with $\bar a=0.7$. This result may be due do the fact that rotating BHs have smaller horizon radii, and thus the
curvature near the horizon is larger; this leads to larger effects in theories, like EdGB, in which the action contains
terms quadratic in the curvature tensor.

\section{Conclusions and outlook}\label{sec:concl}
In this article we have computed the QNMs of a rotating BH in EdGB gravity, a promising example of modified gravity
theory in which the deviations of GR appear in the large-curvature regime. This is the first computation of QNMs of a
rotating BH in modified gravity.

Although we have tested our approach up to order $O({\bar a}^2)$ in the case of a test scalar field, the gravitational
QNMs have only been computed to order $O(\bar a)$. Thus, our results are accurate for $\bar a\ll1$, while for larger
values of the spin (as those typical in actual BHs formed in compact binary coalescences) they should be considered as
an estimate of the actual QNM frequencies, and our calculation should be considered as the first step towards a reliable
computation of the QNMs of rotating BHs in modified gravity theories, to be used in data-analysis frameworks such as
TIGER~\cite{Meidam:2014jpa} or {\sc ParSpec}~\cite{Maselli:2019mjd}.

We compute the shifts of the frequencies due to rotation, $\omega_1$, as defined in Eq.~\eqref{eq:qnms-expansion-2}. In
Eqs.~\eqref{eq:fit} we provide analytical fits of the real and imaginary parts of $\omega_1$, for the fundamental
modes with $l=2,3$.

Our results suggest that the shifts of the frequencies due to general relativity modifications are magnified by rotation;
for $\bar a\sim0.7$, they can be larger than four times the corresponding shifts for a non-rotating BH. 

The next step, presently in preparation~\cite{next}, is the computation of the QNMs to second order in the spin. In this
case, the couplings between perturbations with different values of the harmonic index $l$ can not be neglected (as in
the case of the test scalar field, see Sec.~\ref{sec:scalar2}). Moreover, the gravitational perturbations are coupled
with the scalar perturbation. Thus, both the derivation of the perturbation equations at $O({\bar a}^2)$ and their
numerical implementation are more involved.

\begin{acknowledgments}
We thank Paolo Pani, Emanuele Berti, Andrea Maselli and Ryan McManus for useful suggestions and discussions. The authors
would like to acknowledge networking support by the COST Action CA16104. We also acknowledge support from the Amaldi
Research Center funded by the MIUR programs "Dipartimento di Eccellenza" (CUP: B81I18001170001) and PRIN2017-MB8AEZ.
\end{acknowledgments}

\appendix 

\section{Equations for gravitational perturbations at first order in the spin}\label{app:eq}
The field equations~\eqref{eq:scalar}, \eqref{eq:metric}, linearized in the perturbation around the stationary BH
solution discussed in Sec.~\ref{sec:stationaryebhdgb}, can be written as follows (we follow the same notation
as~\cite{Kojima1992}, and leave implicit the sum over $l,m$):
\begin{align}
  &\left[A^{(I)}_{lm}+ \tilde{A}^{(I)}_{lm} \cos\theta \right]Y^{lm}(\theta) +
  i m  C^{(I)}_{lm} Y^{lm}(\theta) \nonumber\\
  &+ B^{(I)}_{lm} \sin\theta  Y^{lm}_{,\theta}(\theta)=0\,,\label{eq:einst-first-group}
\end{align}
\begin{align}
  &\left[\alpha^{(J)}_{lm}+\tilde{\alpha}^{(J)}_{lm}
    \cos\theta \right]\sin\theta  Y^{lm}_{,\theta}(\theta)- i m \Big[\beta^{(J)}_{lm} \nonumber \\
& + \tilde{\beta}^{(J)}_{lm} \cos\theta  \Big]Y^{lm}(\theta)+{\hat \eta}^{(J)}_{lm} \sin^2\theta Y^{lm}(\theta)   \nonumber\\
  & + {\hat \xi}^{(J)}_{lm} \sin\theta X^{lm}(\theta) + {\hat \gamma}^{(J)}_{lm} sin^2\theta W^{lm}(\theta)
  =0\,,\label{eq:einst-second-group_A}
\end{align}
\begin{align}
  & -\left[\beta^{(J)}_{lm}+\tilde{\beta}^{(J)}_{lm} \cos\theta \right]
  \sin\theta  Y^{lm}_{,\theta}(\theta)- i m \Big[\alpha^{(J)}_{lm} \nonumber \\
 & + \tilde{\alpha}^{(J)}_{lm} \cos\theta \Big]Y^{lm}(\theta) -{\hat \zeta}^{(J)}_{lm} \sin^2\theta  Y^{lm}(\theta) \nonumber\\
  & - {\hat \gamma}^{(J)}_{lm} \sin\theta X^{lm}(\theta) + {\hat \xi}^{(J)}_{lm} \sin^2 \theta W^{lm}(\theta)=0\,,
  \label{eq:einst-second-group_B}
\end{align}
\begin{align}
  & \ \hat f_{lm} \sin\theta  Y^{lm}_{,\theta}(\theta)+i m  \hat g_{lm} Y^{lm}(\theta)
  + \hat s_{lm}  \frac{X^{lm}(\theta)}{\sin\theta} \nonumber \\
& + \hat t_{lm} W^{lm}(\theta)=0\,,\label{eq:einst-third-group_A}
\end{align}
\begin{align}
  &  \hat g_{lm} \sin\theta  Y^{lm}_{,\theta}(\theta)-i m 
  \hat f_{lm} Y^{lm}(\theta) - \hat t_{lm} \frac{X^{lm}(\theta)}{\sin\theta} \nonumber \\
& + \hat s_{lm} W^{lm}(\theta)=0\,,\label{eq:einst-third-group_B}
\end{align}
where in Eq.~\eqref{eq:einst-first-group},
$I=0,1,2,3$ correspond to the components of Einstein's field equations
behaving as scalars under rotations, and $I=4$ corresponds to the scalar field equation; $J=0,1$ in
Eqs.~\eqref{eq:einst-second-group_A}, \eqref{eq:einst-second-group_B} correspond to the components of Einstein's field
equations behaving as vectors under rotations; and Eqs.~\eqref{eq:einst-third-group_A}, \eqref{eq:einst-third-group_B},
correspond to the components of Einstein's field equations behaving as tensors under rotations. We have defined
\begin{align}
& X^{lm}(\theta,\varphi)\equiv 2 Y^{lm}_{,\theta \varphi} - 2 \frac{\cos\theta}{\sin\theta}Y^{lm}_{,\varphi} \\
  & W^{lm}(\theta,\varphi)\equiv -2 \frac{\cos\theta}{\sin\theta}Y_{,\theta} - l(l+1) Y^{lm}
  - 2 \frac{Y^{lm}_{,\varphi\varphi}}{\sin^2 \theta}\,.
\label{eq:XW-usual}
\end{align}
The coefficients $A^{(I)}_{lm}$, $\alpha^{(J)}_{lm}$, $\beta^{(J)}_{lm}$, $\hat s_{lm}$, $\hat t_{lm}$ (of zero-th order
in the spin) and ${\tilde A}^{(I)}_{lm}$, $C^{(I)}_{lm}$, $B^{(I)}_{lm}$, ${\tilde\alpha}^{(J)}_{lm}$,
${\tilde\beta}^{(J)}_{lm}$, ${\hat \eta}^{(J)}_{lm}$, ${\hat \xi}^{(J)}_{lm}$, ${\hat \gamma}^{(J)}_{lm}$, ${\hat
  \zeta}^{(J)}_{lm}$, $\hat f_{lm}$, $\hat g_{lm}$ (of order $O({\bar a})$) are linear combinations of the perturbation
functions $h^{lm}_0(r)$, $h^{lm}_1(r)$, $H^{lm}_0(r)$, $H^{lm}_1(r)$, $H^{lm}_2(r)$, $K^{lm}(r)$, $\Phi^{lm}(r)$ and
their derivatives, with coefficients that depend on $l$ but not on $m$. Their explicit espansions in the coupling
parameter $\zeta$, up to $O(\zeta^6)$, are given in the supplemental {\sc Mathematica} notebook~\cite{notebook}.

We project Eqs.~\eqref{eq:einst-first-group} - \eqref{eq:einst-third-group_B} on the complete set of tensor spherical
harmonics, as in~\cite{Kojima1992}, finding the decoupled equations:
\begin{align}
& A^{(I)}_{{l}m} + i m C^{(I)}_{{l}m} +Q_{l} \left[\tilde{A}^{(I)}_{{l}-1\,m}+({l}-1)B^{(I)}_{{l}-1\,m}\right]\nonumber\\
& +Q_{{l}+1\,m}\left[\tilde{A}^{(I)}_{{l}+1\,m}-({l}+2)B^{(I)}_{{l}+1\,m}\right] =0 
\label{eq:eq1-1st}
\end{align}
\begin{align}
  & {l}({l}+1)\alpha^{(J)}_{{l}m}- i m \left[\tilde{\beta}^{(J)}_{{l}m} + {\hat \zeta}^{(J)}_{{l}m}
    -({l}-1)({l}+2){\hat \xi}^{(J)}_{{lm}}\right]
\nonumber\\
&+ Q_{{l}m} ({l}+1) \left[({l}-1)\tilde{\alpha}^{(J)}_{{l}-1\,m} - {\hat \eta}^{(J)}_{{l}-1\,m}
+({l}-2)({l}-1){\hat \gamma}^{(J)}_{{l}-1\,m}
\right]\nonumber\\
& + Q_{{l}+1} {l} \left[({l}+2)\tilde{\alpha}^{(J)}_{{l}+1\,m} +
    {\hat \eta}^{(J)}_{{l}+1\,m}- ({l}+2)({l}+3){\hat \gamma}^{(J)}_{{l}+1\,m}
\right] = 0
\label{eq:eq2-1st}
\end{align}
\begin{align}
  & {l}({l}+1)\beta^{(J)}_{{l}m} + i m \left[\tilde{\alpha}^{(J)}_{{l}m} + {\hat \eta}^{(J)}_{{l}m} + ({l}-1)({l}+2)
    {\hat \gamma}^{(J)}_{{l}}\right]\nonumber\\
  &  + Q_{{l}m} ({l}+1) \left[({l}-1)\tilde{\beta}^{(J)}_{{l}-1\,m} - {\hat \zeta}^{(J)}_{{l}-1\,m}-({l}-2)({l}-1)
  {\hat \xi}^{(J)}_{{l}-1} \right]\nonumber\\
  & + Q_{{l}+1\,m} {l} \left[({l}+2)\tilde{\beta}^{(J)}_{{l}+1\,m} + {\hat \zeta}^{(J)}_{{l}+1\,m}+ ({l}+2)({l}+3)
  {\hat \xi}^{(J)}_{{l}+1} \right] =0
\label{eq:eq3-1st}
\end{align}
\begin{align}
& {l}({l}-1)({l}+1)({l}+2) \hat s_{lm} - i m ({l}-1)({l}+2)\hat f_{lm}\nonumber \\
  & - Q_{{lm}}({l}-1)({l}+1)({l}+2) \hat g_{{l}-1\,m}\nonumber\\
  &+ Q_{{l}+1}{l}({l}-1)({l}+2)\hat g_{{l}+1\,m} =0
\label{eq:eq4-1st}
\end{align}
\begin{align}
&  {l}({l}-1)({l}+1)({l}+2) \hat t_{lm} + i m ({l}-1)({l}+2)\hat g_{lm} \nonumber\\
  &- Q_{{l}m}({l}-1)({l}+1)({l}+2) \hat f_{{l}-1\,m}\nonumber\\
  &+ Q_{{l}+1\,m}{l}({l}-1)({l}+2)\hat f_{{l}+1\,m} =0\,,\label{eq:eq5-1st}
\end{align}
where $l\ge2$ and
\begin{equation}
Q_{lm}=\sqrt{\frac{(l-m)(l+m)}{(2l-1)(2l+1)}}\,.
\label{eq:defQ}
\end{equation}
As discussed in Sec.~\ref{sec:firstorder}, the QNMs at order $O({\bar a})$ are not affected by the coupling terms, which
can then be neglected. Thus,  Eqs.~\eqref{eq:eq1-1st} - \eqref{eq:eq5-1st} reduce to Eqs.~\eqref{eq:1st-simp}.

\section{Tortoise coordinate for stationary black holes in Einstein-dilaton Gauss-Bonnet gravity}\label{app:tortoise}
The tortoise coordinate $r_*$ is a redefinition of the radial coordinate $r$, mapping the region outside the BH horizon
$r\in[r_{\rm h},+\infty]$ into $r_*\in[-\infty,+\infty]$. By defining
\begin{equation}
  \frac{dr}{dr_*}=F(r)
\end{equation}
the function $F(r)$, safisfying $F(r)\sim r-r_{\rm h}$ for $r\to r_{\rm h}$ and $F(r)\to1$ for $r\to\infty$, can be
found by requiring that the perturbation equations reduce, at the horizon and at infinity, to Eq.~\eqref{eq:expwave}. In
general relativity, $F(r)=1-2M/r$ for Schwarzschild BHs, and $F(r)=(r^2+a^2-2Mr)/(r^2+a^2)\sim(r-r_{\rm h})$ (where
$r_{\rm h}=2M-a^2/(2M)+O(a^4)$) for Kerr BHs. We shall define the tortoise coordinate in EdGB gravity, by requiring the
equation for a test scalar field to have the form~\eqref{eq:expwave}. We have verified (up to first order in the spin)
that this coordinate defines the boundary conditions for the gravitational perturbations as well; this is expected,
since the null ingoing coordinate $v=t+r_*$ (with a similar redefinition of the azimuthal coordinate $\varphi$) should
regularize the coordinate singularity at the horizon, and thus the tortoise coordinate has to be the same for scalar and
gravitational perturbations.

\subsubsection*{Non-rotating BHs}
The metric of a static BH in EdGB gravity is Eq.~\eqref{eq:staticmetric},
\begin{equation}
ds^2=-A(r)dt^2+\frac{dr^2}{B(r)}+r^2d\Omega^2\label{eq:staticmetric2}\,.
\end{equation}
The Klein-Gordon equation for a test scalar field, $\nabla_\mu\nabla^\mu\phi=0$, by expanding
$\phi=\frac{1}{r}\phi^{lm}(r)Y^{lm}(\theta,\varphi)e^{-i\omega t}$, reads (we denote with a prime differentiation with
  respect to $r$):
\begin{align}
  &AB\phi^{lm\prime\prime}+\frac{1}{2}(A'B+B'A)\phi^{lm\prime}\nonumber\\
  &+\left(\omega^2-\frac{A'B+B'A}{2r}-A\frac{l(l+1)}{r^2}\right)\phi^{lm}=0\,.
\label{eq:tort0}
\end{align}
By defining the tortoise coordinate with $F(r)=\sqrt{A(r)B(r)}$~\cite{Salcedo2016}, Eq.~\eqref{eq:tort0} can be written
as
\begin{equation}
  \phi^{lm}_{,r_*r_*}+(\omega^2-V^l)\phi^{lm}=0\label{eq:pha0}
\end{equation}
where $V^l=FF'-Al(l+1)/r^2$, vanishing both at the horizon and at infinity (since $F'\sim r^{-2}$).
Since, for a non-rotating BH, $k_{\rm H}=\omega$, Eq.~\eqref{eq:pha0} coincides with Eq.~\eqref{eq:expwave}.
\subsubsection*{First order in the spin}
At $O(\bar a)$, the metric~\eqref{eq:staticmetric2} acquires the extra term $g_{t\varphi}=-r^2\sin^2\theta\varpi(r)$,
where
\begin{equation}
\varpi(r)=\frac{2J}{r^3}\left[1-\frac{147}{960}\zeta^2\left(1+O\left(\frac{M}{r}\right)\right)+O(\zeta^3)\right]\,.
\end{equation}
Since $g_{\varphi\varphi}=r^2\sin^2\theta$,
\begin{equation}
\Omega_{\rm H}=-\lim_{r\to r_{\rm h}}\frac{g_{t\varphi}}{g_{\varphi\varphi}}=\varpi(r_{\rm h})\,.
\end{equation}
The equation for a test scalar field in this spacetime acquires the extra term $-2m\omega\varpi(r)$. Then, by defining
the tortoise coordinate as in the non-rotating case $F(r)=\sqrt{A(r)B(r)}$, the scalar field equation near the horizon
(neglecting $O({\bar a}^2)$ terms) reads:
\begin{equation}
  \phi^{lm}_{,r_*r_*}+(\omega^2-2m\Omega_{\rm H}\omega)\phi^{lm}=\phi^{lm}_{,r_*r_*}+k_{\rm H}^2\phi^{lm}=0
\end{equation}
in agreement with Eq.~\eqref{eq:expwave}.
\subsubsection*{Second order in the spin}
To define the tortoise coordinate we write $F(r)$ as a generic expansion in powers of $\zeta$, ${\bar a}$ and
$\frac{1}{r}$, such that $F\sim r-r_{\rm h}$ near the horizon and $F(r)\to1$ as $r\to\infty$. By requiring that the
equation for a test scalar field has the form~\eqref{eq:expwave} near the horizon and near infinity, we find
\begin{align}
  F(r)=&\left(1-\frac{r_{\rm h}}{r}\right)\left\{1 -{\bar a}^2\frac{r_{\rm h}(r^2+rr_{\rm h}+r_{\rm h}^2)}{8r^3}\right.\nonumber\\
    &\left.-\zeta^2\left[\frac{r_{\rm h}}{3840 r^4}(147r^3+117r^2r_{\rm h}-526rr_{\rm h}^2+263r_{\rm h}^3)\right.\right.\nonumber\\
    &\left.\left.+{\bar a}^2\frac{r_{\rm h}}{30720 r^3}(375r^2+435rr_{\rm h}+343r_{\rm h}^2)
    \right]\right\}\nonumber\\
  &+O(\zeta^3)+O({\bar a}^3)\,.
\end{align}
In order to obtain this expression we imposed a condition sligthly stronger than Eq.~\eqref{eq:expwave}: we required
that at $r\to\infty$
\[
Z^{lm}_{,r_*r_*}+\omega^2Z^{lm}=\frac{l(l+1)}{r^2}Z^{lm}+O\left(\frac{1}{r^3}\right)\,.
\]
With this further condition, we obtained a better agreement with the results of\,\cite{Cano:2020cao} for the QNMs of a
test scalar field.


\bibliographystyle{utphys}
\bibliography{bibliography}

\end{document}